\documentclass[aps,amsmath,amssymb,superscriptaddress,reprint]{revtex4-2}
\usepackage[utf8]{inputenc}
\usepackage[T1]{fontenc}
\usepackage{etoolbox}
\usepackage{graphicx}
\usepackage{dcolumn}
\usepackage{bm}
\usepackage{blindtext}
\usepackage{physics}
\usepackage{siunitx}
\usepackage{float}
\usepackage{xcolor}
\usepackage{bbold}
\usepackage{relsize}
\usepackage[acronym]{glossaries}
\usepackage[normalem]{ulem}

\newcommand{\One}[0]{\mathlarger{\mathlarger{\mathbb{1}}}}

\newcommand{\ttiny}[1]{\text{\tiny{#1}}}

\newcommand{\MLP}[0]{\text{MLP}}

\definecolor{UniBlue}{HTML}{5F6DA3}
\definecolor{TitleBlue}{HTML}{EBEDF5}
\definecolor{TxtBlue}{HTML}{182D80}
\definecolor{TxtRed}{HTML}{CC3300}
\definecolor{TxtGreen}{HTML}{339900}
\definecolor{SecBlue}{HTML}{434E76}
\definecolor{TxtBlack}{HTML}{212121}
\definecolor{TxtGrey}{HTML}{828282}

\begin{document}

\title{Force-Field-Enhanced Neural Network Interactions: from Local Equivariant Embedding to Atom-in-Molecule properties and long-range effects}
\author{Thomas Plé}
\email[Corresponding author: ]{thomas.ple@sorbonne-université.fr}
\affiliation{Sorbonne Université, LCT, UMR 7616 CNRS, F-75005, Paris, France}
\author{Louis Lagardère}
\email[Corresponding author: ]{louis.lagardere@sorbonne-université.fr}
\affiliation{Sorbonne Université, LCT, UMR 7616 CNRS, F-75005, Paris, France}
\author{Jean-Philip Piquemal}
\email[Corresponding author: ]{jean-philip.piquemal@sorbonne-université.fr}
\affiliation{Sorbonne Université, LCT, UMR 7616 CNRS, F-75005, Paris, France}



\begin{abstract}
We introduce FENNIX (Force-Field-Enhanced Neural Network InteraXions), a hybrid approach between machine-learning and force-fields. We leverage state-of-the-art equivariant neural networks to predict local energy contributions and multiple atom-in-molecule properties that are then used as geometry-dependent parameters for physically-motivated energy terms which account for long-range electrostatics and dispersion. Using high-accuracy \textit{ab initio} data (small organic molecules/dimers), we trained a first version of the model. Exhibiting accurate gas-phase energy predictions, FENNIX is transferable to the condensed phase. It is able to produce stable Molecular Dynamics simulations, including nuclear quantum effects, for water predicting accurate liquid properties. The extrapolating power of the hybrid physically-driven machine learning FENNIX approach is exemplified by computing: i) the solvated alanine dipeptide free energy landscape; ii) the reactive dissociation of small molecules.
\end{abstract}

\maketitle


\section{Introduction}\label{sec:introduction}

In large-scale simulations, interactions between atoms cannot generally be computed from first principles because of the high numerical cost of quantum methods. Instead, they are generally modeled using \textit{force fields} (FFs) that postulate a physically-motivated functional form of the potential energy and are parameterized in order to match \textit{ab initio} energies and/or reproduce experimental data. The most widespread FFs are the so-called \textit{classical} force fields (such as AMBER~\cite{wang2004development} or CHARMM~\cite{vanommeslaeghe2010charmm}) which use a combination of fixed-charge Coulomb potential and Lennard-Jones interactions to model the inter-molecular potential. These models are extremely efficient numerically, allowing the simulation of very large systems over long time scales. Their simple functional form, however, lacks polarization and many-body effects which can be critical to correctly describe some systems (for example solvation in a polar solvent, pi-stacking or complex protein structures~\cite{melcr2019accurate}). More advanced force fields -- such as AMOEBA~\cite{ponder2010current}, TTM~\cite{fanourgakis2008development}, CHARMM Drude~\cite{lemkul2016empirical}, ARROW~\cite{pereyaslavets2022accurate} or SIBFA~\cite{gresh2007anisotropic,naseem2022development} -- have thus been developed in order to explicitly include these effects. These \textit{polarizable} force fields (PFFs) \cite{chapterpol,doi:10.1146/annurev-biophys-070317-033349} are much more flexible and accurate but are significantly costlier. Nonetheless, advances in high-performance computing (HPC), the increase in GPU (Graphical Processing Units) availability and recent methodological developments (advanced iterative solvers) now allow large-scale PFF simulations\cite{adjoua2021tinker}. Both classical and polarizable FFs however assume a fixed connectivity between atoms (\textit{i.e.} covalent bonds cannot be broken), making them unsuitable to study chemical reactions. Some \textit{reactive} force fields -- such as ReaxFF~\cite{van2001reaxff} or Empirical Valence Bond~\cite{warshel1980empirical} -- are actively being developed but are generally specialized towards a relatively narrow class of systems. 

From this brief overview of the domain of force fields, it is clear that a general, many-body reactive model is highly desirable but its design remains an outstanding challenge for the current frameworks. In recent years, considerable attention and resource have been devoted to the development of machine-learning potentials that promise to bridge the accuracy and generality gap between force fields and \textit{ab initio} methods. These models use flexible functional forms from the domain of machine-learning (such as deep neural networks~\cite{shakouri2017accurate,smith2017ani,batzner20223}, body-ordered expansions~\cite{drautz2019atomic,zhu2023mb,yu2022q} or kernel models~\cite{chmiela2018towards,bigi2023wigner}) in order to accurately fit \textit{ab initio} energies, with a numerical cost comparable to standard FFs. A large variety of such models have been developed over the last few years (for example the HD-NNP~\cite{shakouri2017accurate}, ANI~\cite{smith2017ani}, AIMNet~\cite{zubatyuk2019accurate,zubatyuk2021teaching}, DeePMD~\cite{zhang2018deep}, ACE~\cite{drautz2019atomic}, sGDML~\cite{chmiela2018towards}, Tensormol-0.1 \cite{C7SC04934J}, Nequip~\cite{batzner20223}, etc...) and have been applied to small molecular systems~\cite{smith2017ani}, periodic crystals~\cite{faber2015crystal,lysogorskiy2021performant} and more general condensed-phase systems~\cite{zhang2022deep,cheng2019ab}. Among these, the ANI models occupy a particular place as they aim to provide a generic pre-trained potential for a whole class of organic molecules. In this work, we follow a similar strategy.

In order to respect the inherent symmetries of molecular systems, most architectures are designed to be \textit{invariant} with respect to rotations, translations and exchange of identical atoms. These models have shown good accuracy on many systems but usually require large amounts of data to be trained on (of the order of a million molecular configurations). More recently, \textit{equivariant} models (for example Nequip~\cite{batzner20223}, Allegro~\cite{musaelian2022learning}, SpookyNet~\cite{unke2021spookynet}, UNiTE~\cite{qiao2021unite} or GemNet~\cite{gasteiger2021gemnet}) have attracted much attention because of their impressive data efficiency and their ability to generalize more accurately to out-of-distribution configurations~\cite{batzner20223,fu2022forces}.

Most ML models, however, assume a purely local functional form and tend to neglect or implicitly account for long range effects from the training data. The accurate description of long-range interactions is however critical to correctly simulate condensed-phase systems and to describe the structure of large molecular formations (e.g. protein or DNA structure~\cite{gromiha1999importance,york1993effect}). The framework of message-passing neural networks~\cite{gilmer2017neural,haghighatlari2022newtonnet} in principle allows to describe long-range effects by iteratively exchanging information with neighbouring atoms. This approach however have been shown to pose difficulties when applied to large systems as this iterative process is not well suited for parallel architectures because of the associated communications that are required. Furthermore, very long-range effects such as electrostatic interactions would require a large number of iterations (and the molecular graph to be connected) to be captured by a message-passing model, thus imposing a high computational cost~\cite{musaelian2022learning}. Another paradigm to account for long-range effects is the use of global descriptors (for example the Coulomb matrix~\cite{rupp2012fast}) that couple all degrees of freedom without imposing any locality prior. These descriptors are usually used in conjunction with kernel-based models~\cite{dral2017structure,chmiela2018towards} and where shown to be accurate and data-efficient for small to medium-sized molecules and crystals. Although some recent progress have been made to apply these models to larger systems~\cite{chmiela2023accurate}, they cannot tackle systems larger than a few hundreds of atoms due to the $\order{N_{at}^2}$ scaling of the global descriptor size. Some other pure ML multi-scale models are currently being developed (for example the LODE descriptor~\cite{grisafi2019incorporating,grisafi2021multi}) but this area is still in its infancy. On the other hand, quantum perturbation theory (for example Symmetry Adapted Perturbation Theory, SAPT) gives solid grounds to the description of long-range effects in terms of classical electrostatics~\cite{szalewicz2012symmetry} which can be well captured in the FF framework (via multipolar Coulomb interactions and dispersion effects for example~\cite{doi:10.1021/acs.jctc.0c01337}). It thus seems advantageous to combine an ML model -- which excel at predicting short-range properties -- with long-range FF interactions, in order to obtain the best of both approaches. A few models applying this idea have recently been developed (for example HDNNP-Gen4~\cite{ko2021fourth}, PhysNet~\cite{unke2019physnet}, SpookyNet~\cite{unke2021spookynet}, ANIPBE0-MLXDM~\cite{tu2023neural}, q-AQUA-pol~\cite{qu2023interfacing} and others \cite{yang2022transferrable,zhang2022deep,bowman2022delta,zhu2023mb}) and have shown good results across multiple systems. Hybrid FF/ML models thus provide a promising route to more physics-aware ML potentials.

In this paper, we propose a general framework for building force-field-enhanced ML models. We leverage the latest advances in local equivariant neural networks in order to accurately predict short-range energy contributions as well as multiple atom-in-molecule properties that are then used to dynamically parameterize QM-inspired FF energy terms that account for long-range interactions. We show that this architecture allows for highly transferable models that are able to accurately generalize on large molecular systems, as well as in the condensed phase after being trained on small monomers and dimers only. This paper is organized as follows. Section~\ref{sec:nn_architecture} describes the model architecture, from the Allegro~\cite{musaelian2022learning} equivariant embedding to the \textit{output} and \textit{physics} modules. It also describes the particular FF terms that we used for the pretrained model, named FENNIX-OP1, that we provide with this work. Section~\ref{sec:training} focuses on the FENNIX-OP1 model and provides details on its construction, its target properties, the datasets used for training and the different training stages that were required. In section~\ref{sec:model_validation}, we validate the model via several applications. First, we show that the model predicts accurate dissociation energy curves of some simple molecules. We then compute structural properties of liquid water in molecular dynamics simulations including nuclear quantum effects (NQEs) via the recently developed adaptive quantum thermal bath (adQTB)~\cite{mangaud2019fluctuation,mauger2021nuclear}. Indeed, since the model is trained purely on \textit{ab initio} data, the explicit inclusion of NQEs is critical for the correct calculation of thermodynamical properties, as was shown in numerous previous studies~\cite{pereyaslavets2018importance,mauger2022improving}. For this purpose, the adQTB was shown to provide robust approximations of NQEs while being numerically affordable (similar to a classical MD) and thus constitutes a very efficient tool for quickly testing ML models. We then show that the model produces stable dynamics of alanine dipeptide in solution and provides a qualitatively correct description of the torsional free energy profile (computed using an enhanced sampling method\cite{barducci2011metadynamics}); as well as stable dynamics of the 1FSV protein in gas phase. 
Finally, section~\ref{sec:conclusion} provides some conclusions and outlooks for future extensions of the model.

\section{Model Architecture}\label{sec:nn_architecture}
\begin{figure*}
    \centering
    \includegraphics[width=1.\textwidth]{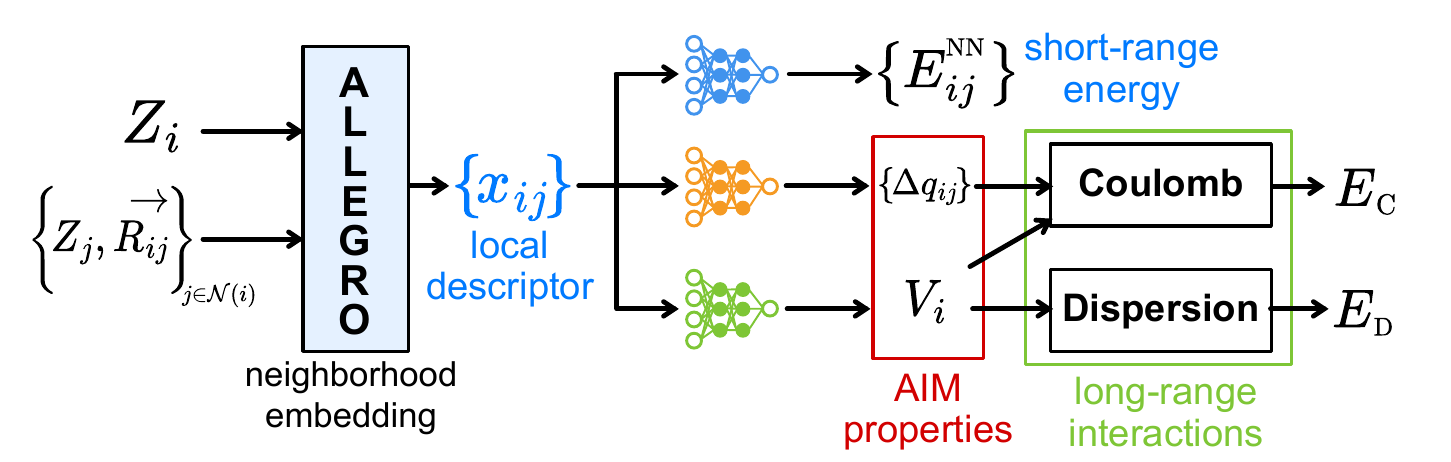}
    \caption{Flow diagram of the FENNIX-OP1 model. }
    \label{fig:fennix_diagram}
\end{figure*}

The FENNIX (Force-field-Enhanced Neural Network Interactions) model is based on a local multi-output equivariant model that processes atomic neighborhoods and predicts multiple atomic or pairwise properties. As an example for this work, we will present a model that outputs local pairwise energy contributions, charges and atomic volumes. The output is subsequently enriched by a "physical" module that computes force field terms such as electrostatic and dispersion energy terms. The core of our model is a slightly modified version of the Allegro local equivariant model presented in ref.~\cite{musaelian2022learning}. We use the Allegro model as a general embedding of atomic pairs which is then fed into independent neural networks that predict the target properties. The choice of the Allegro model is mostly motivated by two of its properties: i) it is strictly local and thus allows for favourable scaling with system size and efficient parallelization; ii) it is equivariant and thus allows the prediction of tensorial properties such as atomic multipoles (which are not used in this work but will be the focus of future improvements of the model). A compact flow diagram of the model is shown in Figure~\ref{fig:fennix_diagram}.

In this section, we will briefly describe the Allegro architecture and our modifications to the model. The theoretical analysis of this architecture was thoroughly done in the original paper~\cite{musaelian2022learning} so that we will only review here the main points necessary to the understanding of the model and our improvements to the architecture. We will then present the output module that processes the Allegro embedding. Finally, we will describe the physical module and the particular functional form for the FENNIX-OP1 potential energy surface that we used in this work.

\subsection{Equivariant embedding using the Allegro architecture}
\subsubsection{Review of the Allegro architecture}
The Allegro model provides a local many-body descriptor $\qty(x_{ij},V_{ij}^{nlp})$ -- interchangeably referred to as embedding in the following -- for each directed pair of atoms with source atom $i$ and destination atom $j$ in the neighborhood $\mathcal{N}(i)$ defined by all the atoms located at a distance shorter than a cutoff radius $r_c$ from atom $i$:
\begin{equation}
    \mathcal{N}(i)=\qty{ k \quad \text{s.t.}\quad \norm{\va{R}_{ik}}<r_c}
\end{equation}
with $\va{R}_{ik}=\va{R}_{k}-\va{R}_{i}$ the vector going from the position of atom $i$ to atom $k$.
The first part of the descriptor $x_{ij}$ is built so that it is invariant under the action of certain geometric symmetries (\textit{i.e.} global rotations, translations and inversions of the system). On the other hand, the second descriptor $V_{ij}^{nlp}$ is composed of features that are equivariant with respect to these symmetries. These features take the form of tensors that are labeled with a channel index $n\in 1,\hdots,N_\ttiny{channels}$, a rotational index $l\in 0,1,\hdots,l_\ttiny{max}$ and a parity index $p\in -1,1$. The rotational index indicates how the tensor transforms under rotation operations: $l=0$ corresponds to scalar/invariant quantities, $l=1$ corresponds to vector-like objects and we refer to $l>=2$ objects as higher-order tensors. The parity index, on the other hand, indicates how the tensor's sign changes under inversion of the coordinate system. The channel index simply allows the model to process multiple features of same $l$ and $p$ indices. In our implementation of the Allegro model, these tensorial objects are handled by the \verb|e3nn| python package~\cite{geiger2022e3nn} which provides high-level classes that represent them and easy-to-use functions to manipulate and combine them while preserving symmetries and global equivariance.

In the following paragraphs, we describe how initial two-body features are computed, how features from neighbors are combined through $N_\ttiny{layers}$ layers of interactions to enrich them with many-body information and how they are filtered at each layer to control the size of the embedding.
~\\

\paragraph*{Initial two-body features}~\\
The Allegro model starts by decomposing each interatomic vector $\{\va{R}_{ij}\}_{j\in\mathcal{N}(i)}$ into fingerprints that are more suitably processed by the network. The interatomic distance $R_{ij}$ is projected onto a radial basis $B(R_{ij})~=~\qty[B_1(R_{ij}),\hdots,B_{N_\ttiny{basis}}(R_{ij})]$ (we use the Bessel basis function with a polynomial envelope~\cite{gasteiger2020directional} that we normalize as in the original paper) and we compute the two-body scalar embedding as:
\begin{equation}
    x_{ij}^\ttiny{2B} = \MLP_\ttiny{2B}\qty[ \One(Z_i)~||~\One(Z_j)~||~B(R_{ij}) ]f_c(R_{ij})
\end{equation}
where $||$ denotes concatenation, $\MLP_{2B}$ is a multilayer perceptron (\textit{i.e.} a fully connected scalar neural network), $f_c(R_{ij})$ is a cutoff function going smoothly to zero as $R_{ij}$ approaches $r_c$ (we use the same polynomial envelope as for the radial basis) and $\One(Z_i)$ (resp. $\One(Z_j)$) is a vector representing the chemical species of the source atom $i$ (resp. destination atom $j$). In the original paper $\One(Z_i)$ is a direct one-hot encoding of the atomic number $Z_i$, meaning that one has to fix in advance the number of species the model will be able to process (the one-hot encoding then defines a simple orthonormal basis which has the same dimensions as the chosen number of species). In section~\ref{sec:positional_encoding}, we propose to modify this one-hot encoding by a positional encoding of coordinates in the periodic table which allows for more flexibility in the treatment of atomic species.

We obtain the two-body equivariant features by projecting the unit vector $\hat{R}_{ij}=\va{R}_{ij}/R_{ij}$ onto a basis of real spherical harmonics $Y_{ij}^{lp}$. We then mix them with radial information with a linear embedding on $N_\ttiny{channels}$ channels:
\begin{equation}
    V_{ij}^{nlp,\ttiny{2B}}=\qty[MLP_\ttiny{embed}^\ttiny{2B}(x_{ij}^\ttiny{2B})]^{nlp}~Y_{ij}^{lp}
\end{equation}
~\\

\paragraph*{Interaction with the local environment}~\\
The two-body embedding $\qty(x_{ij}^\ttiny{2B},V_{ij}^{nlp,\ttiny{2B}})$ is then processed through multiple "interaction" layers that allow to combine information with other atoms in the vicinity of atom $i$. Each interaction layer starts by building a global equivariant neighborhood embedding for atom $i$ from the current scalar embeddings $x_{ik}$ and the spherical harmonics projections $Y_{ik}^{lp}$:
\begin{equation}\label{eq:neighbor_embed}
    \Gamma_i^{nlp,(L)}=\sum_{k\in \mathcal{N}(i)} \qty[MLP_\ttiny{embed}^{(L)}(x_{ik}^{(L-1)})]^{nlp}~Y_{ik}^{lp}
\end{equation}
with $L=1,\hdots,N_\ttiny{layers}$ the layer index and $x_{ik}^{(0)}=x_{ik}^\ttiny{2B}$ and $V_{ij}^{nlp,(0)}=V_{ij}^{nlp,\ttiny{2B}}$. The interaction is then performed via a tensor product of $\Gamma_i^{nlp,(L)}$ with each equivariant embedding $V_{ij}^{nlp,(L-1)}$ (the tensor product is done independently for each channel $n$). The resulting "latent space" 
\begin{equation}
    \mathcal{L}_{ij}^{nmlp,(L)}~=~\qty(\Gamma_i^{nl_1p_1,(L)}\otimes V_{ij}^{nl_2p_2,(L)})^{nmlp}
\end{equation} contains all possible combinations of rotational and parity indices that are allowed by symmetry (\textit{i.e.} such that $\abs{l_1-l_2}\leq l \leq \abs{l_1+l_2}$ and $p=p_1p_2$). Note that since multiple combinations of $(l_1,p_1),(l_2,p_2)$ may produce outputs of indices $(l,p)$, we need to add a multiplicity index $m$ that distinguishes these paths. 
~\\

\paragraph*{Feature filtering and channel mixing}~\\
Finally, the latent space is filtered to obtain the new pairwise embedding. The scalar embedding is combined with the scalar part of the latent space (with every channels and all multiplicities concatenated) to obtain:
\begin{multline}
    x_{ij}^{(L)}=\alpha~x_{ij}^{(L-1)} + \sqrt{1-\alpha^2}~f_c(R_{ij})\\
    \times MLP_\ttiny{latent}^{(L)}\qty[x_{ij}^{(L-1)}~\underset{n,m}{||}~\mathcal{L}_{ij}^{nm01,(L)} ]
\end{multline}
with $0\leq \alpha < 1$ a mixing coefficient that allows to easily propagate scalar information from a layer to the next. Note that the relation between the coefficients ($\alpha$ and $\sqrt{1-\alpha^2}$) is chosen to enforce normalization (see supplementary information of ref.~\cite{musaelian2022learning}). In our implementation, the value of $\alpha$ can be set as a hyperparameter (for example to the value $\alpha=2/\sqrt{5}$ proposed in the original Allegro paper) or can be optimized independently for each layer during the training procedure. 

The new equivariant features are obtained by linearly combining the elements of the latent space with same indices $(l,p)$ from all channels and multiplicities:
\begin{equation}
   V_{ij}^{nlp,(L)}=\sum_{n',m} w^{nlp,(L)}_{n',m}\mathcal{L}_{ij}^{n'mlp,(L)}
\end{equation}
which results in features with the same number of elements as the previous layer. The weights $w^{nlp,(L)}_{n',m}$ are optimized in the training procedure.

The output features of the last layer $\qty(x_{ij}^{(N_\ttiny{layers})},V_{ij}^{nlp,(N_\ttiny{layers})})$ compose the many-body embedding of our model which is passed to the output module to predict the different atomic or pairwise properties that the model is trained on.

\subsubsection{Positional encoding of chemical species} \label{sec:positional_encoding}
While a one-hot encoding allows to represent chemical species in a simple manner, it fixes from the start the number of different species that the model can treat. Thus, if more data becomes available for new species, one would have to retrain the model from scratch in order to accommodate for the new data. More importantly, in such encoding, all species are treated equally and no similarities between species (for example closeness in the periodic table) are provided: the network must learn these correlations purely from data. This encoding is thus suitable when targeting a specific system but might not be the best choice when building a more general chemical model. Some alternatives have been proposed to improve the chemical encoding using for example the ground state electron configuration of each atom~\cite{unke2021spookynet} or a vector mimicking orbitals occupancy~\cite{takamoto2022teanet}.

\begin{figure}
    \centering
    \includegraphics[width=.45\textwidth]{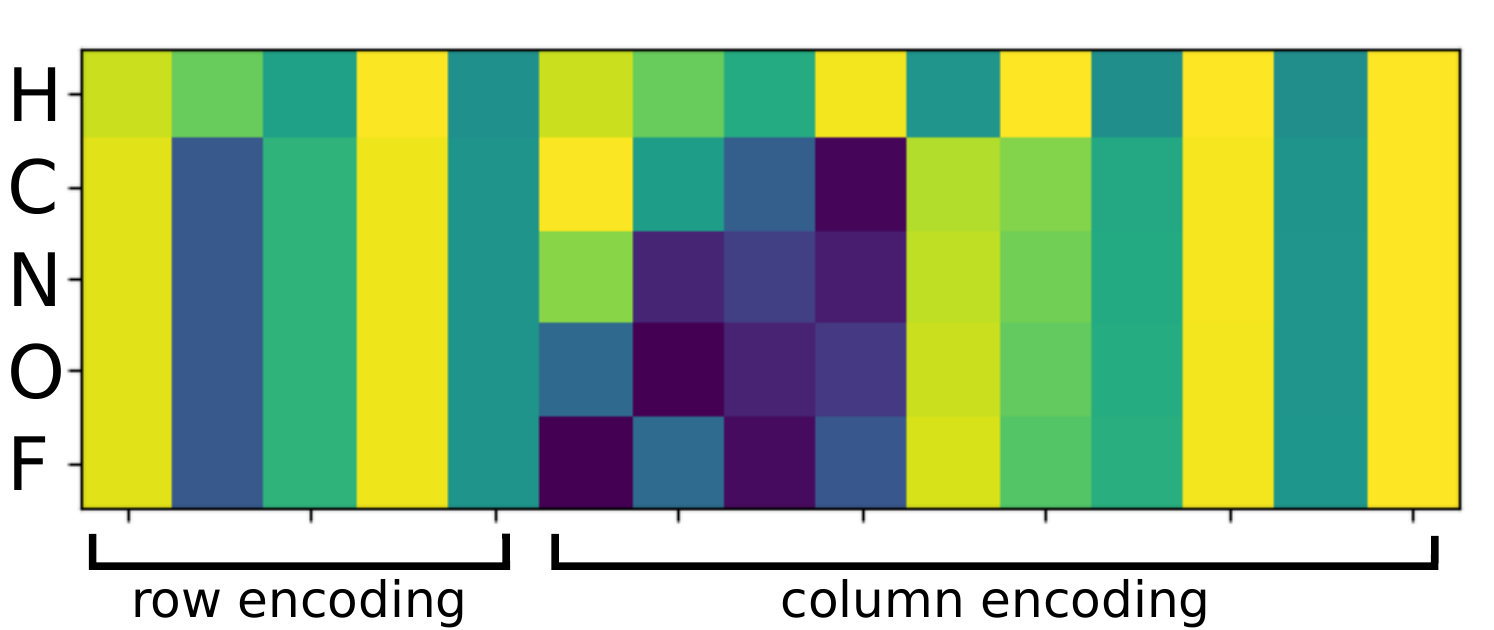}
    \caption{Heatmap of the positional encodings of the chemical species H,C,N,O,F (from top to bottom). The first five columns represent the encoding of the row index in the periodic table, while the last ten columns represent the encoding of the column index.}
    \label{fig:encoding_HCNOS}
\end{figure}

In this work, we propose to use a positional encoding that encodes coordinates in the periodic table using sine and cosine functions of different frequencies. It is inspired from ref.~\cite{vaswani2017attention} where it is used to encode the position of words in sentences in the context of natural language processing.
The column index $c$ is encoded as a vector $e_c$ of dimension $d_\ttiny{col}$ as:
\begin{equation}
    \forall k\in 0,\hdots,d_\ttiny{col},~(e_c)_k=\left\{\begin{aligned}
    &\sin(c/\gamma_\ttiny{col}^{2i/d_{col}}) \quad \text{if k=2i}\\
    &\cos(c/\gamma_\ttiny{col}^{2i/d_{col}}) \quad \text{if k=2i+1}
    \end{aligned}\right.
\end{equation}
and similarly for the row index with dimension $d_\ttiny{row}$ and frequency parameter $\gamma_\ttiny{row}$. For this work, we fix the dimensions and frequency parameters to $d_\ttiny{col}=10$, $d_\ttiny{row}=5$, $\gamma_\ttiny{col}=1000$ and $\gamma_\ttiny{row}=100$ which provide a good compromise between the compactness and richness of the representation. These could also be treated as hyperparameters or even learned during training (for the frequency parameters). The row and column encodings are then concatenated to obtain the full encoding vector $\One(Z_i)=\qty[e_{\ttiny{row}(Z_i)}~||~e_{\ttiny{col}(Z_i)}]$.
Figure~\ref{fig:encoding_HCNOS} shows a heatmap of the positional encoding of the species H,C,N,O and F (from top to bottom). We see that the first five columns are the same for all the heavy atoms as they represent the row encoding (the second row of the periodic table in this case) while they are different from the first line corresponding to the Hydrogen. We also see that the last ten columns are different for all the species shown here as they are all on different columns of the periodic table. The motivation behind using this positional encoding is that we hypothesized that having similar encodings for species sharing a row or a column might help with generalization and allow to transfer learned knowledge from a species to another, thus requiring less training data. Interestingly, the positional embedding defines a smooth function of the coordinates in the periodic table (of which integer coordinates are actual chemical elements) and allows to "interpolate" between chemical species. Furthermore, as stated in ref.~\cite{vaswani2017attention} the encoding for index $c+k$ can be represented as a linear function of the encoding for index $c$, which might further help with inferring similarities. Additional features such as the ionization state could be encoded in the same manner (though we restrict ourselves to neutral atoms in this work, thus only requiring the knowledge of chemical species).

\subsection{Output module}
\label{sec:output_module}
After computing the embedding from the Allegro model, we use it as input for independent MLPs for each target property. In the following, we will simply denote $\qty(x_{ij},V_{ij}^{nlp})$ the output from the last Allegro layer (thus dropping the $(L)$ layer index). The current implementation also allows some modularity in the composition of inputs and on the operations done on the outputs. For example, the input can exploit either the scalar embedding to obtain invariant properties via a standard MLP as
\begin{equation}
    o_{ij}=MLP_\ttiny{out}[x_{ij}]
\end{equation}
, or both the scalar and tensorial embeddings via a linear projection of $V_{ij}^{nlp}$
\begin{equation}
    O_{ij}^{mlp}=\sum_n \qty[MLP_\ttiny{out}(x_{ij})]^{mlp}_n~V_{ij}^{nlp}
\end{equation}
. For atom-wise properties, the pairwise outputs are simply summed up on the central atom. For properties that should sum up to zero (for example partial atomic charges), the outputs $o_{ij}$ and $o_{ji}$ can be antisymmetrized (which for partial charges is equivalent to charge exchange between neighbouring atom pairs).
Furthermore, in order to impose constraints on invariant outputs, a final activation function can optionally be applied. This is for example useful when the output targets a positive quantity (for instance an atomic volume) for which we can apply a \textit{softplus} function, a probability for which we may apply a sigmoid function or a discrete probability distribution (in the case of multidimensional $o_{ij}$) for which a \textit{softmax} function can be used.

Further modifications can optionally be applied. For instance, the input can be filtered according to an additional shorter-range cutoff for example one distinguishing between bonded and non-bonded pairs. Finally, the two-body embedding $x_{ij}^\ttiny{2B}$ can be used in place of or concatenated to (as it is done in the FENNIX-OP1 model) the final embedding to use as input. This allows the output MLP to easily access simple pairwise information and should let the Allegro embedding specialize in finer many-body correlations. This compositional approach allows for a great flexibility in the model's output, which is especially useful when experimenting with a ML parametrization of physical models.

The output module for the FENNIX-OP1 model is composed of three targets: a local energy contribution $E^\ttiny{NN}_{i}=\sum_{j\in \mathcal{N}(i)} E^\ttiny{NN}_{ij}$ (which is a simple scalar output), an atomic partial charge $q_i^\ttiny{NN}=\sum_{j\in \mathcal{N}(i)} \Delta q_{ij} - \Delta q_{ji}$ (through antisymmetrized charge exchange) and an atomic volume $v_i^\ttiny{NN}$ (constrained to be positive).

\subsection{Physics module and energy functional form}\label{sec:force_field_module}
Finally, the physics module uses the results from the output module and feed them into physically-motivated models to enrich the output. 

In the case of FENNIX-OP1, the force field module is composed of an electrostatic energy term $E^\ttiny{C}_{ij}$ and a pairwise dispersion term $E^\ttiny{D}_{ij}$. The functional form of FENNIX-OP1 is then given by:
\begin{equation}
\label{eq:total_energy}
    E^\ttiny{OP1} = \sum_{i} E^\ttiny{NN}_{i} 
    +\sum_{i,j<i} E^\ttiny{C}_{ij}
    +\sum_{i,j<i} E^\ttiny{D}_{ij}
\end{equation}
The first term $E^\ttiny{NN}_{i}$ is the neural network contribution that accounts for short-range interactions that we introduced in section~\ref{sec:output_module}. We model the electrostatic interaction $E^\ttiny{C}_{ij}$ via a Coulomb potential with fluctuating charges and the Piquemal charge penetration model~\cite{piquemal2003improved}:
\begin{multline}
    E^\ttiny{C}_{ij} = \frac{1}{R_{ij}}\Big[N_iN_j
    + N_j(q_i-N_i)f_\alpha(R_{ij}/r_i^\ttiny{vdw}) \\ + N_i(q_j-N_j)f_\alpha(R_{ij}/r_j^\ttiny{vdw})\\
    + (q_i-N_i)(q_j-N_j)f_\beta(R_{ij}/r_i^\ttiny{vdw})f_\beta(R_{ij}/r_j^\ttiny{vdw})
    \Big]
\end{multline}
where $N_i$ is the number of valence electrons of atom $i$ 
, $q_i=\epsilon~q^\ttiny{NN}_i$ is the environment-dependent charge of atom $i$ predicted by the neural network (with an adjustable universal scaling parameter $\epsilon$) and $f_\alpha(r)=1-e^{-\alpha r}$ and $f_\beta(r)=1-e^{-\beta r}$ are damping functions with $\alpha$ and $\beta$ adjustable parameters that are assumed to be universal and where
\begin{equation}
   r^\ttiny{vdw}_i=\qty(\frac{v^\ttiny{NN}_i}{v^\ttiny{free}_i})^\frac{1}{3}r^\ttiny{vdw,free}_i
\end{equation}
is the environment-dependent van der Waals radius of atom $i$ with $v^\ttiny{NN}_i/v^\ttiny{free}_i$ the atomic volume ratio predicted by the neural network.

Finally, the dispersion interaction $E^\ttiny{D}_{ij}$ is computed using the pairwise Tkatchenko-Scheffler model~\cite{tkatchenko2009accurate}:
\begin{equation}
    E^\ttiny{D}_{ij}=-\frac{C_\ttiny{6,ij}}{R_{ij}^6}\sigma_{ij}(R_{ij})
\end{equation}
with the combination rule:
\begin{equation}
    C_{6,ij}=\frac{2C_{6,i}C_{6,j}}
        {C_{6,i}\frac{\alpha_{j}}{\alpha_{i}} + {C_{6,j}\frac{\alpha_{i}}{\alpha_{j}}}}
\end{equation}
and the environment-dependent homonuclear parameters:
\begin{equation}
    C_{6,i}=\qty(\frac{v^\ttiny{NN}_i}{v^\ttiny{free}_i})^2 C_{6,i}^\ttiny{free}\qquad;\qquad
    \alpha_{i}=\qty(\frac{v^\ttiny{NN}_i}{v^\ttiny{free}_i})\alpha_{i}^\ttiny{free}
\end{equation}
with $\alpha_i^\ttiny{free}$ the isolated atom polarizabilities,  $C_{6,i}^\ttiny{free}$ the isolated atom sixth order dispersion coefficients and the sigmoid damping function:
\begin{equation}
    \sigma_{ij}(R_{ij})=\qty[1+e^{-\gamma\qty(\frac{1}{s}\frac{R_{ij}}{r^\ttiny{vdw}_i + r^\ttiny{vdw}_j}-1)}]^{-1}
\end{equation}
with $\gamma$ and $s$ adjustable parameters that we assume to be universal.
~\\

We furthermore mention that the physics module is not limited in principle to compute energy terms. The aim of this module is to be a general physical interface between the ML embedding and the final target properties. For example, we use it in the FENNIX-OP1 model to correct for ML-predicted charge exchanges that would deplete the valence shell of an atom. The implementation also provides charge exchange via a simple bond-capacity model~\cite{poier2019describing} that leverages ML-predicted atom-in-molecule electronegativities and which capabilities will be explored in future iterations of the FENNIX model. Each physical model is implemented as a simple Pytorch Module that takes as input a dictionary which contains previously computed properties. It then adds its contribution and outputs the enriched dictionary. These physical submodules can then easily be chained, and it facilitates the implementation of additional models.

\section{Datasets and training procedure}~\label{sec:training}
In this section, we start by providing some details on the construction of FENNIX-OP1 model. We then review the datasets that we used for training the model and its target properties. Finally, we will focus on the non-trivial task of training a multi-output FENNIX model.

\subsection{FENNIX-OP1 model architecture}

In the FENNIX-OP1, chemical species are encoded using the positional encoding defined in section~\ref{sec:positional_encoding}, with 5 dimensions for the row encoding and 10 dimensions for the column encoding. We use $N_\ttiny{basis}=10$ Bessel basis functions for the radial embedding, with a cutoff distance $r_c=5.2$ \AA~ and a smoothing parameter $p=3$ for the polynomial envelope. We used 256 features for the scalar embedding and 10 channels with a maximum rotational index $l_\ttiny{max}=2$ for all equivariant features. The two-body scalar MLP ($\MLP_\ttiny{2B}$) is composed of two hidden layers with 64 and 128 neurons respectively with SiLU activation function. The two-body embedding MLP ($MLP_\ttiny{embed}^\ttiny{2B}$) for the initial equivariant features is a simple linear projection of the two-body embedding with no activation function. The embedding is constructed using 3 Allegro layers. In each layer, the embedding MLP ($MLP_\ttiny{embed}^{(L)}$) is a simple linear projection with no activation function and the latent MLP ($MLP_\ttiny{latent}^{(L)}$) contains two hidden layers both with 256 neurons and SiLU activation function. The mixing coefficient $\alpha$ is initially set to the original $2/\sqrt{5}$ and is optimized independently for each layer during training.
~\\

The output module is composed of three independent identical MLPs for the short-range energy contribution, the charge exchange and the atomic volume. The input of these MLPs is the concatenation of the two-body scalar embedding and the final scalar embedding. They have 5 hidden layers with 256, 128, 64, 32 and 16 neurons. In total, the model has approximately 1.2 million parameters.

The output module also has a "constant" submodule that simply provides the atomic reference energies. Importantly, we used the CCSD(T) isolated atom energies as a reference instead of the average atomic energy over the dataset that is typically advised when training a ML model. Indeed, this reference energy has a physical meaning -- which is the energy of an atom when it has no neighbors to interact with -- and is not solely a convenience parameter for facilitating the training. In particular, this choice has important consequences for the description of molecular dissociation~\cite{batatia2022design}, as will become clear in section~\ref{sec:bond_dissociation}. 
~\\

Finally, the physics module is composed of four submodules: a charge correction module, a charge scaling module, the Coulomb interaction module and the dispersion module. These last two modules implement the physical interactions using the corresponding equations described in section~\ref{sec:force_field_module}. The charge scaling module simply scales all charges by a constant factor in order to compensate for the lack of higher-order multipoles in the permanent electrostatics.
The charge correction module antisymmetrizes the charge exchanges and ensures that atoms do not unphysically deplete their valence shell. Indeed, if we assume that only valence electrons can be transferred, an atom cannot have a partial charge larger than $+N_i$ (which is particularly important for the charge penetration model that we use). This constraint is not ensured by the neural network model so we need to enforce it \textit{a posteriori}. The constraint is achieved by transferring back some electrons from neighbouring atoms that drained too much charge. First the unphysical "hole" in the valence shell is computed using a smooth function (to ensure smooth gradients of the final coulomb energy) on the basis that an atom cannot lose more than 95 percent of its valence electrons. Charges that sum up to the valence hole are then transferred from neighbouring atoms that took charges, proportional to the quantity drained. This procedure should enforce the constraint in a single iteration for most non-pathological cases. Reassuringly however, the charge correction is almost never needed after training (\textit{i.e.} the valence hole is almost always zero), even in condensed-phase MD simulations for which the model was not explicitly trained.

\subsection{Datasets and target properties}
For the FENNIX-OP1 model, we chose to reproduce high-level coupled-cluster energies and we thus selected three of the few freely-available and generalist datasets that provide such data: the ANI-1ccx~\cite{smith2017ani} dataset, the DES370K~\cite{donchev2021quantum} dataset and the q-AQUA dimers dataset~\cite{yu2022q}.

The \textbf{ANI-1ccx dataset} provides approximately 500,000 CCSD(T)/CBS total energies of various neutral monomers and dimers (composed of the elements H,C,N,O) in equilibrium and out-of-equilibrium configurations. It also provides atomic forces at DFT level that were crucial to speed-up the beginning of the training procedure. Importantly, it provides partial charges and atomic volumes obtained from a Minimal Basis Iterative Stockholder~\cite{verstraelen2016minimal} (MBIS) partitioning of the DFT electronic density that were used to parametrize the force field terms.

The \textbf{DES370K dataset} is composed of approximately 370,000 CCSD(T)/CBS interaction energies of diverse dimers in various configurations. It comprises both neutral and charged molecules. The latter were discarded for the training of the FENNIX-OP1 model as out-of-scope for this study. It also provides SAPT decompositions of the interaction energies that we used to regularize the Coulomb interactions. 

The \textbf{q-AQUA dimers dataset} is composed of more than 70000 water dimer interaction energies at CCSD(T)/CBS level. We used this dataset in the end of the training in order to fine-tune the model for handling water.
~\\

The FENNIX-OP1 model has then three target properties that use the available data: the CCSD(T) total energy of the system as modelled by $V_\ttiny{OP1}$ described in section~\ref{sec:force_field_module}, the MBIS partial charges $q_i^\ttiny{NN}$ and the ratio of MBIS volumes to free atom volumes $v_i^\ttiny{NN}/v_i^\ttiny{free}$.

\subsection{Training procedure}
The training of a multi-output force-field-enhanced neural network revealed to be a non-trivial task. Indeed, the inter-dependencies between target properties (for example partial charges and the electrostatic contribution to the total energy) seemed to pose difficulties for the standard optimization methods, which implied that a brute-force optimization of the whole model would not give satisfactory results. To overcome this difficulty, we resorted to a training procedure in multiple stages that is described in the following of this section.

Furthermore, in order to obtain a final model that was stable when performing molecular dynamics, we found that strong regularization was needed, both in the form of standard weight decay and also physically-motivated loss contributions, as detailed later. Throughout, we used the AdamW optimizer~\cite{loshchilov2017decoupled} implemented in Pytorch, as its algorithm for weight decay provides one of the best compromise between training speed and accuracy. We used a fairly strong weight decay parameter of 0.5 for all the parameters in the output module and no weight decay for the Allegro parameters. In all stages, we used the same random 10\% of the dataset as a validation set and optimized the model using mini-batches of 256 configurations from the ANI-1ccx and 64 configurations from DES370K and q-AQUA when they are used.
~\\

The training procedure for the FENNIX-OP1 model required four stages.
In the first stage, the model was trained to reproduce DFT total energies and forces using the short-range contribution $V^\ttiny{NN}$ only. We also trained at the same time the charges and atomic volumes to reproduce the MBIS targets. At this stage, only the ANI-1ccx dataset was used. The loss function for this stage is given  by:
\begin{multline}
    L^{(1)}=\lambda_E \qty(E^\ttiny{DFT}-E^\ttiny{NN})^2 + \lambda_F \sum_{j=1}^{3}\sum_{i=1}^{N_{at}}\qty(F^\ttiny{DFT}_{ij} - F^\ttiny{NN}_{ij})^2\\
    +\lambda_q\sum_{i=1}^{N_{at}}\qty(q^\ttiny{MBIS}_i-q^{NN}_i)^2+\lambda_v\sum_{i=1}^{N_{at}}\qty(\frac{v^\ttiny{MBIS}_i}{v^\ttiny{free}_i}-\frac{v^\ttiny{NN}_i}{v^\ttiny{free}_i})^2
\end{multline}
with $\lambda_E=0.001$, $\lambda_F=1$ and $\lambda_q=\lambda_v=1000$ and $F^\ttiny{NN}$ is the model's predicted force obtained by automatic differentiation (Pytorch's autograd) of the total energy $E^\ttiny{NN}$. As suggested in previous studies~\cite{musaelian2022learning}, we strongly favour learning forces over energies in the beginning to accelerate the training procedure. We note that the provided loss is for a single configuration and that, in practice, it is averaged over the configurations in the mini-batch. We further added a regularization in order to minimize the off-diagonal elements of the covariance matrix of the scalar embedding's features over a batch. This promotes learning statistically independent features in the embedding which should be favourable for a multi-output network and we found that it led to models with better generalization capabilities.
In this stage, we train all the parameters in the the embedding and output modules with a starting learning rate of $10^{-3}$. Furthermore, we used a learning rate scheduler that reduces the learning rate when the error on the training set stops diminishing for a few steps (we set the patience of the scheduler to 10 epochs and the learning rate scaling factor to 0.8). After about 100 epochs, progress of both training and validation steps slowed down and we modified the energy and force parameters to $\lambda_E=0.01$ and $\lambda_F=0.1$ in order to obtain a more balanced training. We stopped the first stage when the learning rate reached $10^{-4}$. 

In the second stage, we freeze the embedding parameters and output MLPs for charge and volumes and activate the Coulomb and dispersion energy terms. We then retrain the short-range energy MLP so that the full $V^\ttiny{OP1}$ of eq.~\eqref{eq:total_energy} reproduces DFT energies and forces. The loss function for this stage is:
\begin{equation}
    L^{(2)}=\lambda_E \qty(E^\ttiny{DFT}-E^\ttiny{OP1})^2 + \lambda_F \sum_{j=1}^{3}\sum_{i=1}^{N_{at}}\qty(F^\ttiny{DFT}_{ij} - F^\ttiny{OP1}_{ij})^2
\end{equation}
with the same weights as in the end of the previous stage. Freezing the embedding ensures that the predicted volumes and charges are not modified in this training stage. Since the energy target is modified, the errors starts much higher than at the end of the previous stage and quickly decreased. When the error on the train and validation sets drop to the same order as in the previous stage, the full model is unfrozen and training is resumed until the error stops decreasing. 

In the third stage, the embedding and charge and volumes MLPs are frozen again and the energy MLP is finally retrained to reproduce CCSD(T) energies from both ANI-1ccx total energies and DES370K interaction energies. We used the same type of mean-square loss functions with $\lambda_E=0.1$ and $\lambda_\ttiny{DES}=5$. Again, we train the energy MLP until the error is close to the previous stage, unfreeze the whole model and optimize again all the parameters. For this stage, we also optimize the parameters from the physical module in order to reach the lowest error possible. We stop the training when the learning rate reaches $10^{-5}$.

In the last training stage, we refine the model for water by including the q-AQUA water dimers interaction energies in the loss function. We also generated batches of randomly deformed water monomers (with very large deformations) and trained the model to reproduce forces from the highly accurate Partridge-Schwenke potential energy surface (that was itself fitted on high-accuracy coupled-cluster data). This was particularly helpful to reproduce the molecule's bending energy surface away from equilibrium where fewer data are available in the ANI-1ccx dataset.
~\\

At the end of the training procedure, the model reached a root mean square error (RMSE) of less that 4 kcal/mol for CCSD(T) total energies on both validation and training sets of the ANI-1ccx dataset. While it was possible to reach much lower errors with less regularized models (less than 1 kcal/mol), we found that they were unstable when performing MD and concluded that they were overfitting the data. We thus favoured a more strongly regularized, perhaps slightly underfitted model that allowed for better generalization in the condensed phase. The model also reached a RMSE of about 0.35 kcal/mol on both DES370K and q-AQUA datasets. Finally, it reached a RMSE of about 0.017 e for charges and about 0.017 for volume ratios. As for total energies, less regularized models allowed for much lower errors (less than 0.008 e for charges for example) but with visible overfitting, leading to irregular coulomb interactions and unstable dynamics.

As a validation for the water interactions, we used the standard energy benchmarks when training force fields for water. The model gives a RMSE of 0.13 kcal/mol on the Smith dimer set~\cite{smith1990transition} which are representative of the most stable water dimer configurations. This low error is not surprising as the Smith dimers are very close to configurations present in the q-AQUA dataset on which the model was trained. For a more challenging test, we used a set of typical water clusters up to hexamers. For these, the model achieved a RMSE lower than 2 kcal/mol, which is comparable to our recent Q-AMOEBA model~\cite{mauger2022improving}  which was specifically trained to reproduce these energies. In the next section, we investigate more thoroughly the validity, transferability and robustness of the model up to the unforgiving test of condensed-phase molecular dynamics including nuclear quantum effects.

\section{Model validation}~\label{sec:model_validation}

In this section, we validate the FENNIX-OP1 model using a few examples of applications. First, we compute bond dissociation energy profiles of typical small molecules and show that the model is able to consistently break covalent bonds. Then, we show that the model is able to produce stable and accurate MD simulations of condensed-phase water including nuclear quantum effects. For this study, we included nuclear quantum effects using the adaptive quantum thermal bath (adQTB) method introduced in ref.~\cite{mangaud2019fluctuation}. We showed in previous studies~\cite{mauger2021nuclear} that the adQTB provides an efficient and accurate alternative to path integrals -- the gold standard method for including NQEs in MD simulations -- at a cost similar to classical MD. For these two examples, we compare our results to the ANI models~\cite{smith2017ani} that have a comparable scope as FENNIX-OP1 in terms of chemical diversity and were trained on similar datasets (note that we used the ANI models as provided in the torchANI package and did not re-train them). Finally, we show that FENNIX-OP1 is able to produce stable dynamics of organic molecules solvated in water and provides a good qualitative description of the torsional free energy landscape of the alanine dipeptide in solution.
~\\

All calculations for this work were performed on an Nvidia A100 GPU using our custom TorchNFF package that is built on top of Pytorch~\cite{pytorchpaper} and provides the implementation for FENNIX as well as a simple MD algorithm for NVT simulations in periodic boundary conditions (with Ewald summation for electrostatic interactions) and an efficient Pytorch implementation of the adQTB. TorchNFF is in an early development version and is thus not yet optimized. For example, a FENNIX-OP1 simulations of a box of 216 molecules of water in periodic boundary conditions can currently reach about 0.8 ns/day of adQTB simulation (with a timestep of 0.5 fs) on a single A100 GPU. 

\subsection{Bond dissociation energy profiles}
~\label{sec:bond_dissociation}
\begin{figure*}
    \centering
    \includegraphics[width=0.9\textwidth]{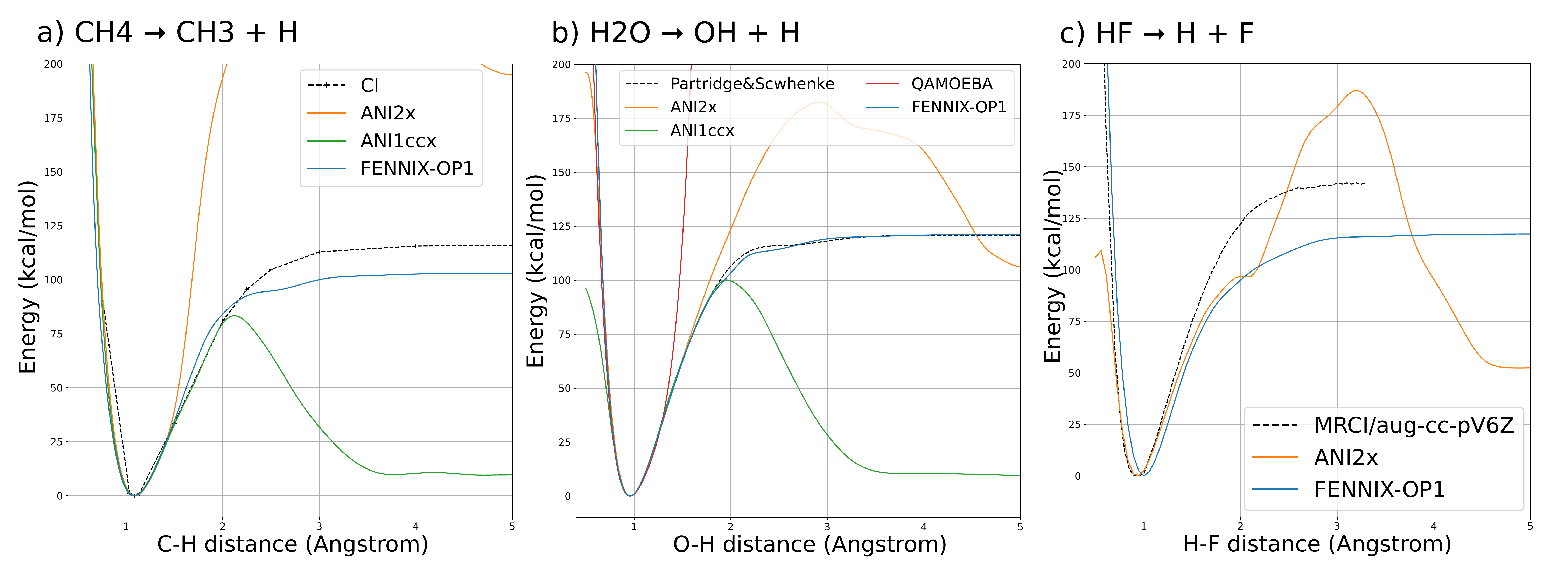}
    \caption{Potential energy curves for the dissociations of: a) methane CH4$\rightarrow$CH3+H with the CH3 group fixed at the CH4 equilibrium geometry; b) water H2O$\rightarrow$OH+H with OH and angle fixed at the H2O equilibrium geometry; c) HF$\rightarrow$H+F  }
    \label{fig:dissociation_curves}
\end{figure*}
We first validate the training and the choice of reference energy on potential energy curves for the dissociation of covalent bonds in small molecules. This kind of bond breaking is a fundamental step in the process of many chemical reactions, for example in enzyme-catalyzed reactions~\cite{warshel2006electrostatic}, and is key to the description of mass spectroscopy experiments~\cite{baer1996unimolecular,de2007mass}. They are however difficult to accurately model: force fields usually forbid them by design (by assigning a fixed connectivity) and \textit{ab initio} approaches require the use of expensive multi-reference calculations to correctly describe the dissociation process. Their practical description thus usually requires the use of specifically designed force fields (such as ReaxFF) or low-dimensional potential energy surfaces.

Figure~\ref{fig:dissociation_curves} shows the potential energy curves for a) the dissociation of a hydrogen atom in methane; b) the asymmetric dissociation of the water molecule; c) the dissociation of the H-F molecule computed with FENNIX-OP1 and compared with reference quantum calculations from the literature (multi-reference CI/6-31G** from ref~\cite{hirst1985ab} for CH4, Partridge-Scwhenke PES~\cite{partridge1997determination} that was fitted on CCSD(T) data and multi-reference CI/aug-cc-pV6Z from ref~\cite{duohui2014mrci} for the H-F molecule) and the ANI models. We see that FENNIX-OP1 consistently captures a smooth dissociation curve thanks to the physically-motivated choice of reference energy. On the other hand, the ANI models, for which the reference energy was set according to average values over the dataset, fail to reproduce the dissociation curves for large distances.
FENNIX-OP1 agrees particularly well with the reference for water as the Partridge-Schwenke PES was included in the training set. For the other two molecules, it tends to underestimate the dissociation barrier. Interestingly, while the training data did not contain covalent interaction energies for F, the model is still able to reasonably generalize its prediction. In appendix~\ref{appendix:dissociation}, we describe how the model learned a generic representation of dissociation energy profiles, independently of chemical species. The knowledge of the chemical species involved in the bond (via the chemical encoding) then provides the details of the energy curve such as the equilibrium bond length and the height of the dissociation barrier, with some transferability across the periodic table thanks to the positional encoding introduced in section~\ref{sec:positional_encoding}. 


\subsection{Structural and spectroscopic properties of liquid water}\label{sec:liquid_water}
\begin{figure}
    \centering
    \includegraphics[width=0.47\textwidth]{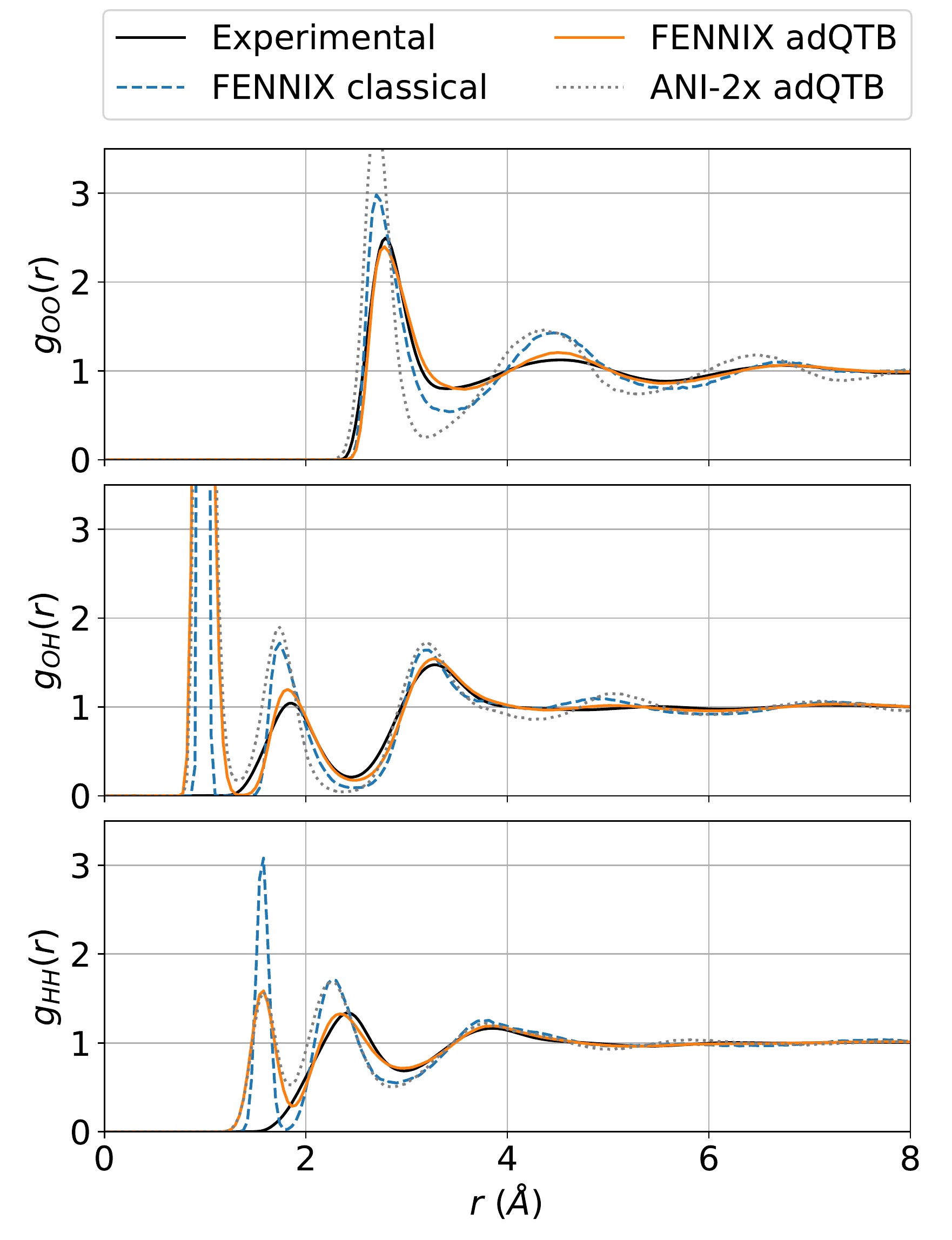}
    \caption{Partial radial distribution functions of liquid water at 300K simulated using classical MD and adQTB MD with the FENNIX-OP1 model, compared to experimental results from~\cite{soper2013radial} and ANI-2x (adQTB) results.}
    \label{fig:gr_water}
\end{figure}
In order to test the robustness of the model, we performed molecular dynamics simulations of liquid water. Since the model was fitted purely on \textit{ab initio} data, it was critical to explicitly include nuclear quantum effects in order to accurately compute thermodynamical properties~\cite{pereyaslavets2018importance}. We used the adaptive quantum thermal bath method to include NQEs~\cite{mangaud2019fluctuation}, as it was previously shown to be a robust alternative to the more costly path integrals MD~\cite{mauger2021nuclear}.
All simulations were performed for a box of 216 molecules in the NVT ensemble at 300K and experimental density, with the BAOAB integrator~\cite{Leimkuhler2012} and a timestep of 0.5 fs. Coulomb interactions in periodic boundary conditions were handled using the Ewald summation method~\cite{ewald1921ewald}, that we directly implemented using PyTorch operations so that we can leverage its automatic differentiation capabilities to obtain atomic forces. We equilibrated the system for 100 ps, which was sufficient to thermalize the system and converge the adQTB parameters. We then computed the radial distribution functions (RDFs) from 1 ns of MD simulation. Figure~\ref{fig:gr_water} shows the partial RDFs of water simulated using adQTB and classical MD with the FENNIX-OP1 model compared to experimental results from~\cite{soper2013radial}. As a baseline, we also compare to ANI-2x results with adQTB MD. 
We see that FENNIX-OP1 is able to accurately capture the subtle structure of liquid water and agrees well with experimental results when nuclear quantum effects are included. This is quite remarkable as the model was not explicitly trained on condensed phase reference data. 
On the other hand, ANI-2x predicts a largely over-structured liquid, which was shown to be mainly due to the insufficient quality of 
 the DFT reference used for training the model~\cite{rosenberger2021modeling}. The ANI-1ccx model, that was trained with CCSD(T) reference data, produced very accurate radial distribution functions at the beginning of the simulations, thus validating the necessity of accurate CCSD(T) reference data to be able to describe liquid water. It was however not stable for simulation times longer than a few tens of picoseconds (even in classical MD with a smaller 0.1 fs timestep). As pointed out in ref.~\cite{rosenberger2021modeling}, this can be affected to the lack of long-range effects that are important to maintain the structure of the liquid. Furthermore, we show in Supplementary Information, that neglecting the long-range tail of the dispersion interactions in FENNIX-OP1 leads to visible distortions of the radial distribution functions. The sensitivity of the structural results even to these relatively small interactions (compared to the larger Coulomb interactions) reinforces the necessity to accurately capture long-range interactions with physics-informed models.

\begin{figure}
    \centering
    \includegraphics[width=0.47\textwidth]{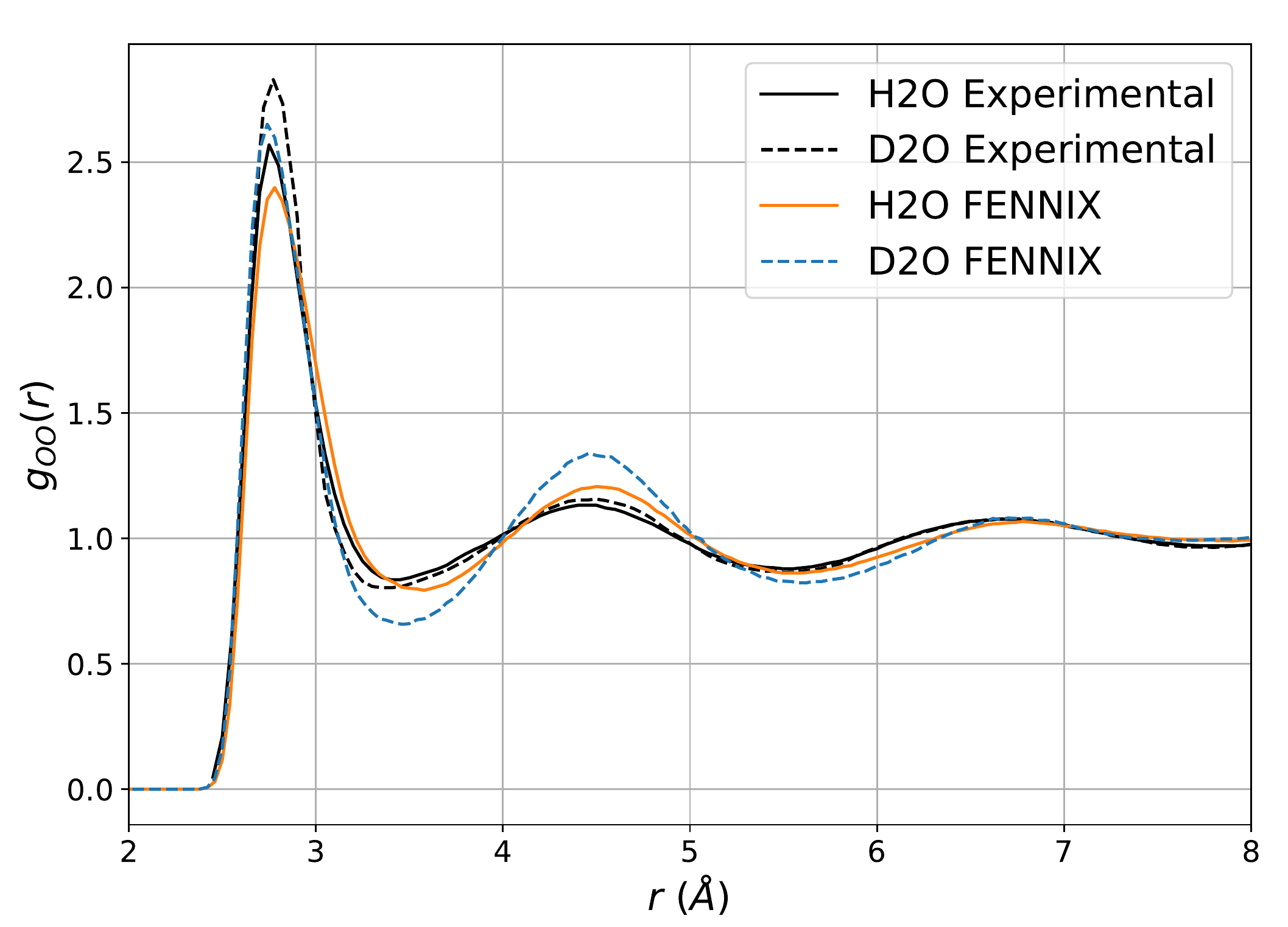}
    \caption{O-O radial distribution computed using adQTB and the FENNIX-OP1 model for light and heavy water, compared to experimental results from ref.~\cite{soper2008quantum}.}
    \label{fig:gr_heavy_water}
\end{figure}

We then explored the impact of deuteration on the structural properties. Such thermodynamical isotope effects arise purely from NQEs (since the classical Boltzmann distribution does not depend on the mass of the atoms) and can be used to experimentally probe the impact of quantum effects on chemical processes. In liquid water, deuteration tends to strengthen the liquid's structure, resulting in slightly sharper peaks in the O-O radial distribution function~\cite{soper2008quantum,soper2013radial,ceriotti2016nuclear}. Reproducing this effect is a stringent test for the interaction model as it requires to accurately capture the subtle balance of competing quantum effects that affect the hydrogen bonds~\cite{li2011quantum,ceriotti2016nuclear}, which is challenging even for accurate DFT functionals~\cite{li2022static}.
We showed in earlier works that, provided an accurate potential energy surface, the adQTB is able to reliably capture these isotope effects~\cite{mauger2021nuclear,mauger2022improving}. Figure~\ref{fig:gr_heavy_water} shows the O-O radial distribution function of light and heavy water at 300K computed using FENNIX-OP1 and compared to experimental results from ref.~\cite{soper2008quantum}. The model captures the strengthening of the structure with deuteration but seems to amplify the effect, indicating slightly too weak hydrogen bonds~\cite{li2011quantum,ceriotti2016nuclear,mauger2022improving}.

Additionally, FENNIX-OP1 predicts an enthalpy of vaporization for light water around 13 kcal/mol, which is slightly overestimated compared to the experimental result of 10.51 kcal/mol~\cite{wagner1993international}. This indicates a tendency of the model to form too strong hydrogen bonds in the condensed phase.  We expect that training the model using interaction energies from larger molecular clusters would improve the results overall. Indeed, it was recently shown with the q-AQUA(-pol)~\cite{yu2022q,qu2023interfacing} and MB-pol(2023)~\cite{zhu2023mb} water models that including interaction energies of water clusters containing at least up to four molecules in the training set was necessary to obtain state-of-the-art results on both gas-phase and condensed-phase properties.

Contrary to ML models that only predict interaction energies, FENNIX provides environment-dependent atomic charges, enabling us to estimate infrared spectra through the computation of time correlation functions of the (nonlinear) total dipole moment from the adQTB dynamics. The results, shown in Supporting Information, are in qualitative agreement with experimental data for the main features, albeit a 150 cm$^{-1}$ red-shift of the bending peak (similar to the q-SPC/Fw model~\cite{paesani2006accurate}) and a broadening of low-frequency features. Comparison to spectra computed using classical MD shows that the model captures the typical red-shift of the stretching peak due to nuclear quantum effects~\cite{habershon2009competing,benson2020quantum,ple2021anharmonic}. 
Still, better accuracy on liquid water's IR spectrum are obtained by models specifically targeting water, for example in Ref.~\cite{liu2015quantum}, that include a refined many-body dipole moment surface. 
Furthermore, the FENNIX-OP1 spectra lack typical features produced by the slow dynamics of induced dipole~\cite{impey1982spectroscopic,habershon2009competing,mauger2022improving}. Even though FENNIX-OP1 is able to capture local polarization effects via fluctuating charges, this subtle many-body effect cannot be reproduced without an explicit treatment of long-range polarization, as is done for example in (Q-)AMOEBA~\cite{ren2003polarizable,mauger2022improving}, SIBFA~\cite{naseem2022development} or ARROW~\cite{pereyaslavets2022accurate}. Future iterations of FENNIX will then focus on refining the physical model for including such interactions and better reproducing the system's dipole moment and its evolution in a framework including multipolar electrostatics. 

\subsection{Enhanced sampling of the torsional free energy profile of solvated alanine dipetide}

Lastly, we performed molecular dynamics simulations (using enhanced sampling) of alanine dipetide explicitly solvated in a cubic box of water of length 30 \AA~(with periodic boundary conditions). The alanine dipeptide is a fundamental building block for larger biomolecules and its accurate description is thus critical. This system is also a typical benchmark for both interaction models and enhanced sampling methods and was recently shown to be extremely challenging for ML potentials~\cite{fu2022forces}. It thus constitutes an interesting test case for our potential. 

\begin{figure}
    \centering
    \includegraphics[width=0.5\textwidth]{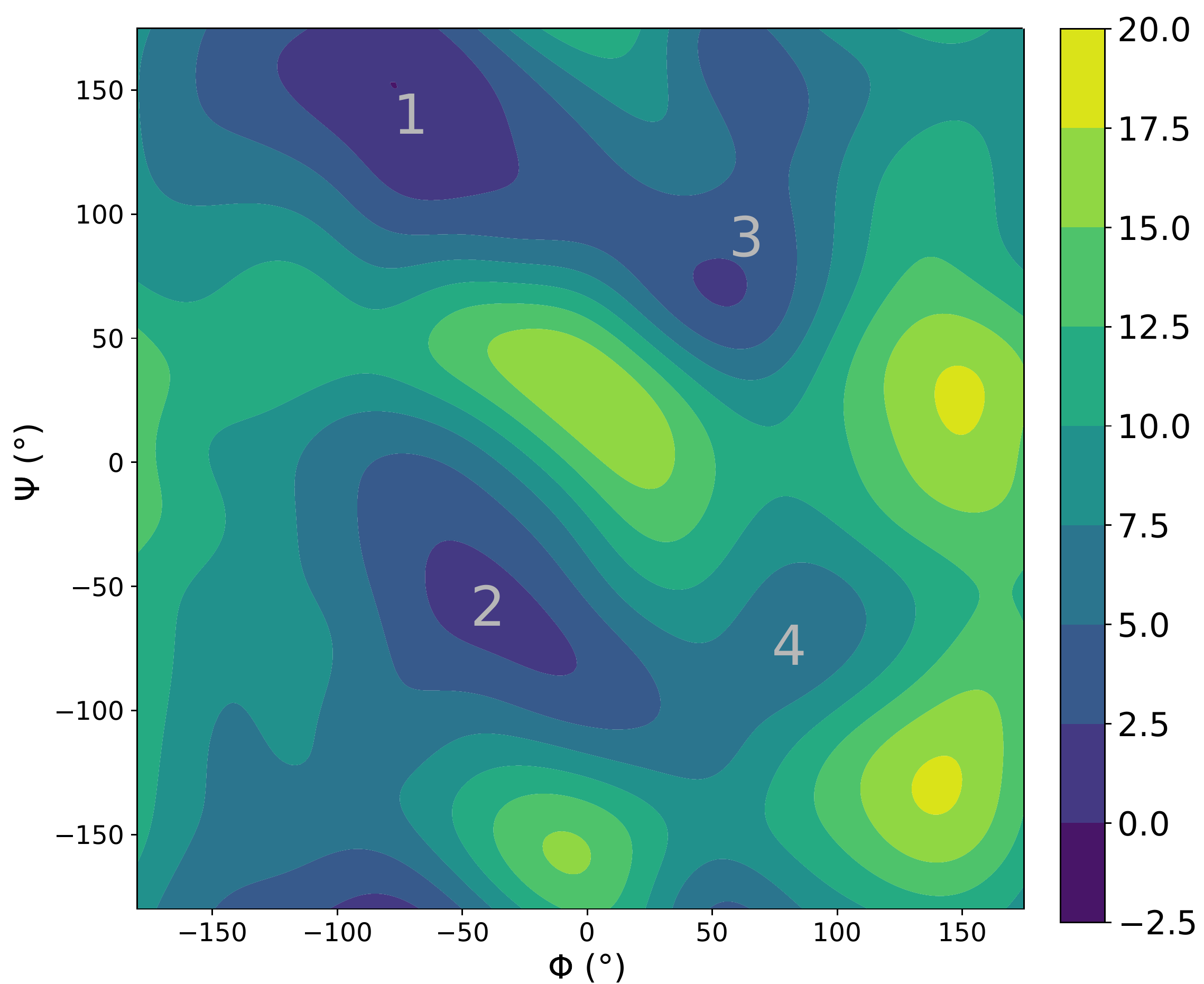}
    \caption{Contour plot of the free energy profile of the two torsional angles of solvated alanine dipeptide computed using FENNIX-OP1 model. Sampling was enhanced using well-tempered metadynamics for the two torsional angles. Energies in kcal/mol.}
    \label{fig:alanine}
\end{figure}

Figure~\ref{fig:alanine} shows the joint free energy profile, obtained after 3 ns of classical well-tempered metadynamics~\cite{barducci2008well} simulation (performed using the PLUMED library~\cite{plumed2019promoting}) with the FENNIX-OP1 model, for the two dihedral angles defining the torsional degrees of freedom of the molecule. The model correctly assigns most of the probability to the two expected energy basins denoted 1 and 2 on figure~\ref{fig:alanine}. It was also able to explore the less probable states denoted 3 and 4 in the figure. The model however seems to underestimate the barrier at $\Phi=0$, thus allowing too many transitions between states 1 and 2 to 3 and 4. We note that this simulation was performed using classical MD, as the adQTB is not directly compatible with enhanced sampling methods. It would be interesting to explore the impact of nuclear quantum effects on this energy profile. Indeed, as we showed above, nuclear quantum effects drastically modify the structure of water. Since the torsional free energy profile is strongly affected by the solvent (see the difference between solvated and gas phase alanine dipeptide for example in ref.~\cite{smith1999alanine}), we can expect important changes when going from classical to quantum dynamics. This calculation would require long and costly path-integrals simulations or the development of adequate enhanced sampling techniques for the adQTB, that we leave for future works.

\subsection{Perspective: gas-phase simulation of a protein}
\begin{figure}
    \centering
    \includegraphics[width=0.5\textwidth]{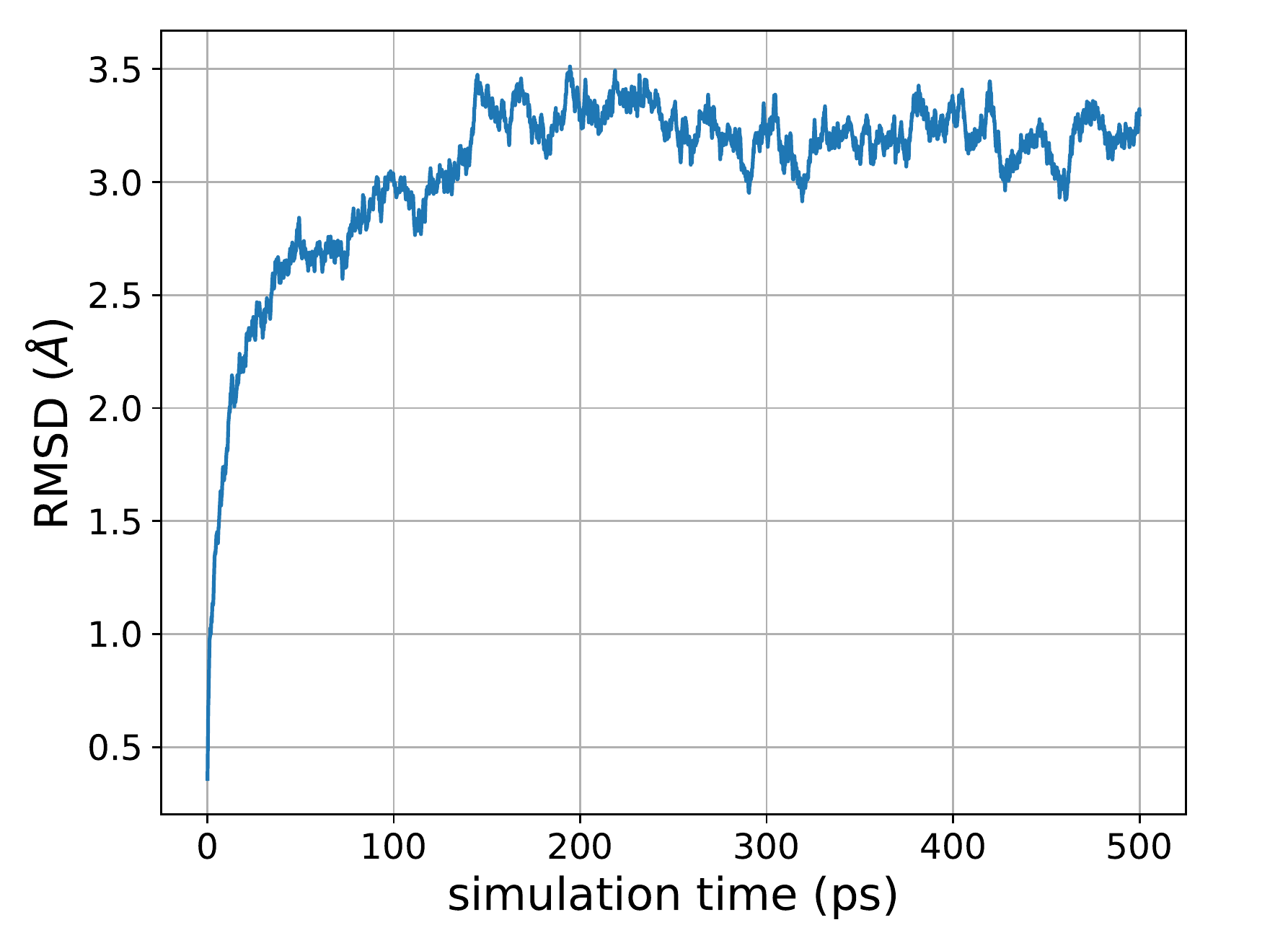}
    \caption{Root-mean-square deviation of the 1FSV protein with respect to the PDB structure during a 500 ps adQTB MD simulation in gas phase with the FENNIX-OP1 model.}
    \label{fig:rmsd_protein}
\end{figure}

As a perspective, we pushed the model towards larger molecular structures that are not included in the training data. We performed molecular dynamics, including nuclear quantum effects, of the 26-residue 1FSV protein~\cite{dahiyat1997novo} in gas phase. It contains around 500 atoms with a few charged groups. Since the model is not designed to handle ionic species, we first performed a simulation where we neutralized each charged group (by addition or subtraction of a proton). We ran an adQTB MD simulation for 500 ps with a timestep of 0.5 fs. Starting from the PDB structure, the dynamics was stable and the protein folded to a more compact form typical of the gas phase. The RMSD of the backbone with respect to the PDB stucture stabilized after around 150 ps of simulation to a value of $\sim$3.2 \AA, as shown in figure~\ref{fig:rmsd_protein}. This fast structural rearrangement is mostly driven by the long-range interactions present in the model whose magnitude is greater in gas-phase compared to condensed-phase due to a lack of screening by the solvent. These preliminary results illustrate the robustness of the force-field-enhanced ML approach, able to generalize to much larger and complex systems than that included in the training set.

Even though the model was not trained on charged species, we ran a short MD simulation of the protein in solution by explicitly distributing the ionic charges among the atoms of the charged groups. After a few picoseconds, however, the simulation displayed instabilities due to unphysical proton transfers around the charged groups and too large Coulomb interactions that collapsed the molecule.
In order to produce quantitative results and stable dynamics in this context, next iterations of the FENNIX model will need to explicitly handle charged systems, which will require a more thorough dataset. The recently introduced SPICE dataset~\cite{eastman2023spice} could be a good complement to the ones already used in this work as it provides reference data for many charged molecules and focuses on biological systems (however at the lower quality DFT level). Furthermore, it will require finer description of molecular interactions through the inclusion of more advanced energy terms (such as multipolar electrostatics and polarization).

\section{Conclusion and outlooks}\label{sec:conclusion}
We introduced FENNIX, a general class of models representing molecular interactions using an hybrid ML/FF approach. It builds upon the latest equivariant architectures to construct an embedding of the local chemical environment which is fed into a multi-output module predicting atom-in-molecule properties. The latter allows for a flexible design of potential energy terms. 
We apply this strategy to design a specific model: FENNIX-OP1, capturing both short-range interactions, with a dedicated ML energy term, and long-range ones using predicted atomic volumes and charges that parametrize both a charge penetration corrected electrostatic energy and a Tkatchenko-Scheffler like dispersion one. FENNIX-OP1 is first trained on both the ANI-1ccx and the DES370K datasets which contain monomers, dimers and a few multimeric structures focusing on small neutral organic molecules. It is then refined for water using the q-AQUA dimers dataset and the highly accurate Partridge-Schwenke potential energy surface for a final RMSE of 0.35 kcal/mol on interaction energies.
We then showed that the model is stable during molecular dynamics (including NQEs via adQTB) and able to generalize to condensed phase (despite the lack of reference data), capturing structural properties of bulk water and solvated organic molecules. More precisely, it qualitatively reproduces the torsional free energy profile of the solvated alanine dipeptide using enhanced sampling techniques and yields stable trajectories of the 1FSV protein in gas phase. 
Interestingly, we showed that the model learned some generic concepts (such as dissociation profiles) independently of the chemical species, which are then specifically refined for each element -- with some transferability across the periodic table -- thanks to the continuous encoding of their positions in the periodic table. The good accuracy of the model for bond dissociations in simple molecules is a promising first step toward the general description of more complex chemical reactions. 

All in all, this work shows the extrapolating power of hybrid physically-driven machine learning approaches and paves the way to even more refined models using enriched data sets (in particular for charged systems) and additional energy terms in the spirit of advanced polarizable force-fields such as SIBFA that include multipolar electrostatics and many-body polarization and dispersion. These will enable the study of reactive complex systems such as proton transfers between DNA pairs or enzyme catalyzed reactions where such an accurate description of molecular interactions is mandatory. An important question going forward is to what extent general and reactive models can compete with specialized potentials. Indeed, specialized models~\cite{yu2022q,qu2023interfacing,zhu2023mb} can achieve very good accuracy on multiple properties of the system they target, that general ones (such as FENNIX-OP1) currently do not match. On the other hand, the flexibility of general models enables the simulation of more complex systems involving a large variety of constituents, as is often the case in biochemistry. A promising strategy that could provide a solid middle-ground is the fine-tuning approach where a general model is slightly re-trained with limited data to better represent a specific target system. Further work will focus on the development and use of more general datasets (such as the SPICE dataset~\cite{eastman2023spice}), the extension of the FENNIX approach and its inclusion in the optimized Deep-HP platform\cite{inizan2022scalable,ple2022routine} present within the Tinker-HP package\cite{C7SC04531J, adjoua2021tinker}.



\section*{Author contributions statement}

T. P. performed simulations and contributed new code;\\
T. P., L. L, J--P. P contributed new methodology; \\
T. P., L. L, J--P. P  contributed analytical tool; \\ 
T. P., L. L, J--P. P analyzed data.\\ 
T. P., L. L, J--P. P wrote the paper;\\
J--P. P designed the research. 
\section*{Acknowledgements}
This work was made possible thanks to funding from the European Research Council (ERC) under the European Union's Horizon 2020 research and innovation program (grant agreement No 810367), project EMC2. Computations have been performed at GENCI (IDRIS, Orsay, France and TGCC, Bruyères le Chatel) on grant no A0130712052. 


\section*{Supporting Information available}
The supporting information contains the results for infrared spectra of liquid water computed using the FENNIX-OP1 model.
\appendix

\section{Dissociations across chemical space}\label{appendix:dissociation}
\begin{figure}[t!]
    \centering
    \includegraphics[width=0.4\textwidth]{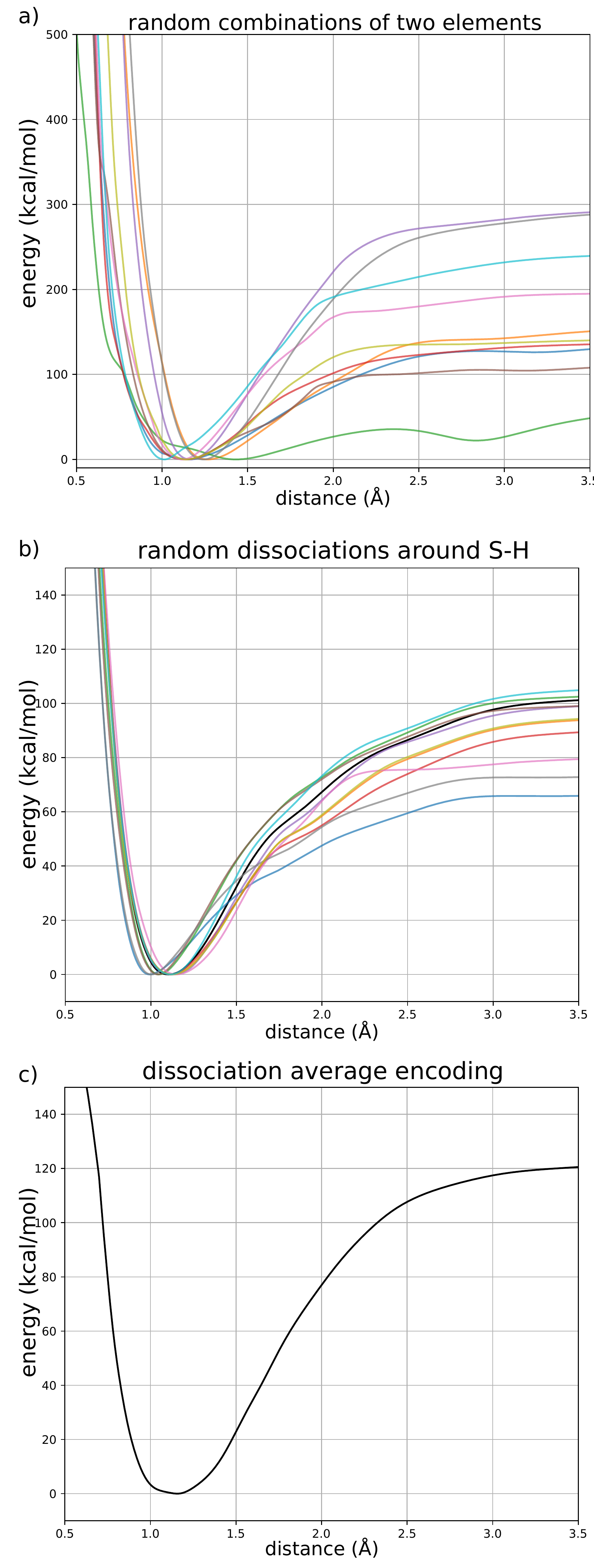}
    \caption{a) dissociation curves for random couples of chemical element (from H to Xe); b) S-H dissociation (in black) and dissociation curves obtained by perturbing the periodic table coordinates of S by a Gaussian noise (of mean 0 and variance 1); c) generic homonuclear dissociation curve from the encoding obtained by averaging the encodings from H to Xe.}
    \label{fig:random_dissociations}
\end{figure}
The goal of this section is to illustrate that FENNIX-OP1 learned a generic representation of dissociation energy profiles (independently of the chemical species involved) and that the positional encoding of chemical species provides the specific properties of each bond (for example the equilibrium distance and bond dissociation energy). To this end, we present three numerical experiments shown in Figure~\ref{fig:random_dissociations}: a) dissociation curves for random couples of chemical element; b) S-H dissociation with the periodic table coordinates of S perturbed by Gaussian noise; c) generic dissociation curve for the average encoding.

Figure~\ref{fig:random_dissociations}.a) shows dissociation curves for ten random couples of chemical elements (from H to Xe). Even if most of of these elements were not included in the training data, FENNIX-OP1 predicts meaningful energy profiles with a clear minimum and short-range repulsion, which shows that the general shape of the dissociation curve is not tied to the chemical encoding.

We then exploited the continuity of the chemical encoding with respect to coordinates in the periodic table in order to show its impact on a specific dissociation curve. Figure~\ref{fig:random_dissociations}.b) show the dissociation profile of S-H (in black) as well as other profiles obtained by shifting the coordinates of S in the periodic table by a two-dimensional Gaussian random vector (with mean 0 and standard deviation 1). We see that moving this way in the periodic table produces similarly-shaped energy profiles but with different bond parameters (bond length, dissociation energy, frequency at equilibrium...) which are dictated by the specific value of the encoding.

\begin{figure}[ht!]
    \centering
    \includegraphics[width=0.5\textwidth]{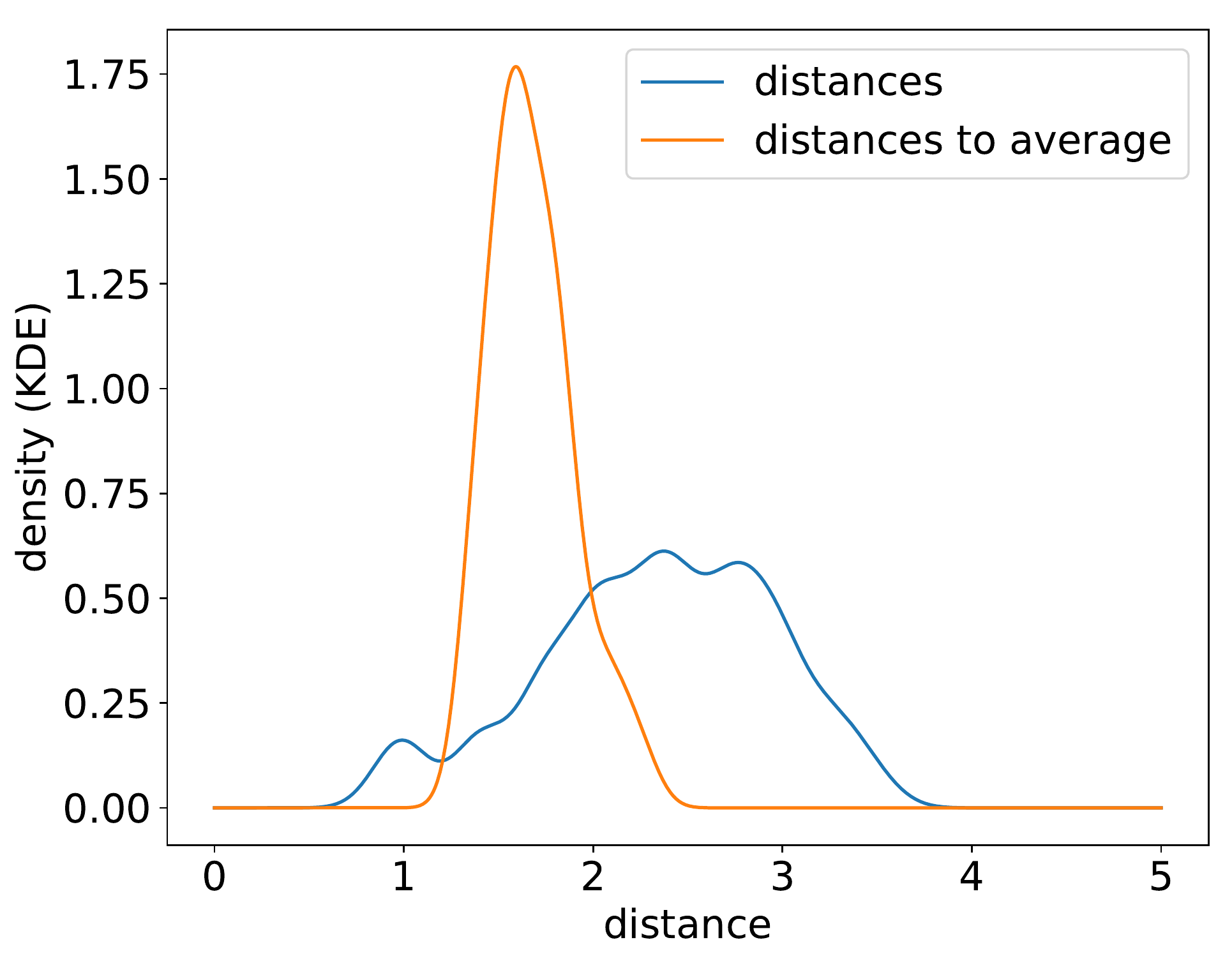}
    \caption{Distribution of euclidean distances in the chemical encoding space (smoothed using a Gaussian kernel density estimation) between true chemical elements to each other (from all combinations of pairs of elements) (blue) and between true chemical elements to the average encoding (orange).}
    \label{fig:distance_distribution_encoding}
\end{figure}

To uncover the generic energy profile learned by the model, we show in Figure~\ref{fig:random_dissociations}.c) the homonuclear dissociation energy profile of a fictitious element which encoding is the average encoding of elements from H to Xe. In the encoding space, this average element is roughly at equal distance from all the other elements, and closer in average to all elements than true elements are to each other, as can be seen from the distributions of distances in the encoding space shown in Figure.~\ref{fig:distance_distribution_encoding}. The dissociation curve shown in Figure~\ref{fig:random_dissociations}.c) can thus be thought of as the prototypical energy profile learned by the model, independently of the chemical species.

\newpage
\bibliography{biblio}
\end{document}